\documentstyle[aps]{revtex} 
\begin{document}
\twocolumn
\wideabs{
\title{Tolman wormholes violate the strong energy condition}
\author{
David Hochberg$^{+}$, 
Carmen Molina--Par\'{\i}s$^{++}$, 
and  Matt Visser$^{+++}$
}
\address{
$^{+}$Laboratorio de Astrof\'{\i}sica Espacial y F\'{\i}sica Fundamental,
Apartado 50727, 28080 Madrid, Spain\\
$^{++}$Theoretical Division, 
Los Alamos National Laboratory,
Los Alamos, New Mexico 87545, USA\\
$^{+++}$Physics Department, Washington University,
Saint Louis, Missouri 63130-4899, USA
}
\date{8 October 1998; \LaTeX-ed \today}
\maketitle

{\small

For an arbitrary Tolman wormhole, unconstrained by symmetry, we shall
define the bounce in terms of a $3$--dimensional edgeless
achronal spacelike hypersurface of minimal volume.  (Zero trace for
the extrinsic curvature plus a ``flare--out'' condition.)  This
enables us to severely constrain the geometry of spacetime at and near
the bounce and to derive general theorems regarding violations of the
energy conditions---theorems that do not involve geodesic averaging
but nevertheless apply to situations much more general than the highly
symmetric FRW--based subclass of Tolman wormholes.  [For example: even
under the mildest of hypotheses, the strong energy condition (SEC)
must be violated.] Alternatively, one can dispense with the minimal
volume condition and define a generic bounce entirely in terms of the
motion of test particles (future-pointing timelike geodesics), by
looking at the expansion of their timelike geodesic congruences.  One
re-confirms that the SEC must be violated at or near the bounce. In
contrast, it is easy to arrange for all the other standard energy
conditions to be satisfied.

}


} 



\section{Introduction}
\def\tr{\hbox{\rm tr}}
\def\implies{\Rightarrow}
\def\conv{\hbox{\rm conv}}
\def\Re{ {\cal R} }
\def\epsfbox#1{}

A so-called Tolman wormhole is formed if a collapsing universe somehow
halts its contraction before encountering a big crunch singularity and
then re-expands. Thus Tolman wormholes are prototypes for modeling the
``oscillating universe'' cosmologies that were in vogue in the
1930's~\cite{Einstein,Tolman}.  In many cases, the precise nature of
the ``bounce'' that was invoked to drive re-expansion was left
unspecified (singular cusp? angular momentum barrier? analytic
extension through the singularity?). In this article we shall
explicitly assume that the ``bounce'' occurs at a moment when the
geometry is non-singular and shall seek to extract as much generic
information as possible about constraints that can then be placed on
the bounce.

Specifically, we shall assume that the universe reaches a moment of
minimum spatial volume, and call this minimum volume edgeless achronal
spacelike hypersurface the ``bounce''. The Tolman wormhole will then
be taken to be some suitable open region of spacetime surrounding this
bounce.  If we additionally assume rotational and translational
symmetry, then the case of the corresponding bouncing
Friedman--Robertson--Walker (FRW) universe has already been considered
in \cite{Bounce}. We shall use that Letter as guidance, but in this
article wish to avoid unnecessary symmetry constraints, and so shall
also seek guidance from recent analyses of generic traversable
wormholes~\cite{Morris-Thorne,MTY,Visser} and their throats
\cite{HV:1,HV:2,HV:3,HV:4}.

Tolman wormholes \cite{Bounce} and traversable
wormholes~\cite{Morris-Thorne,MTY,Visser} are rather different
objects: the Tolman wormhole is intrinsically time dependent and
involves a ``bounce'' for the entire universe, so the throat is a
$3$--dimensional spacelike hypersurface [timelike normal], whereas
traversable wormholes are local
objects~\cite{Morris-Thorne,MTY,Visser} whose throats are
$(2+1)$--dimensional timelike hypersurfaces [spacelike normals].
Nevertheless, we shall see that many parts of the analysis can be
naturally carried over from one case to the other.

The 1988 analysis of Morris and Thorne revitalized interest in {\em
traversable} wormholes~\cite{Morris-Thorne} when they were able to
show that traversable wormholes were compatible with our current
understanding of general relativity and semiclassical quantum
gravity---but that there was a definite price to be paid---one had to
admit violations of the null energy condition (NEC).  More precisely,
what Morris and Thorne showed was equivalent to the statement that for
static spherically symmetric traversable wormholes there must be an
open region surrounding the throat over which the NEC is
violated~\cite{Morris-Thorne,MTY,Visser}. For spherically symmetric
homogeneous Tolman wormholes (bouncing FRW universes) the analogous
statement is that there is an open temporal region surrounding the
bounce on which the SEC must be violated~\cite{Bounce}. (Traversable
wormholes are cosmologically interesting in their own
right~\cite{Roman,Hochberg-Kephart}, but we will not directly address
that topic in this paper.)

To set up the analysis for a generic Tolman wormhole, we first have to
define exactly what we mean by a such a wormhole---we find that there
is a nice {\em geometrical\,} (not topological) characterization of
the existence of, and location of, the ``bounce''. This
characterization is developed in terms of a hypersurface of minimal
area, subject to a ``flare--out'' condition that generalizes that
of~\cite{Bounce}.  With this definition in place, we can develop a
number of theorems about SEC violations at or near the bounce.  While
SEC violations at or near the bounce are unavoidable, it is
relatively easy to satisfy {\em all} the other standard energy
conditions.

We develop a general analysis of energy condition
violations in Tolman wormholes. (This analysis is based largely
on~\cite{Bounce,HV:1,HV:2,HV:3,HV:4}. For an analysis using similar
techniques applied to static vacuum and electrovac black holes see
Israel~\cite{Israel:67,Israel:68}. A related decomposition applied to
the collapse problem is addressed in~\cite{Israel:86}.) In view of the
preceding discussion we want to get away from the notion that topology
is the intrinsic defining feature of wormholes, either traversable or
Tolman, and instead focus on the geometry of the wormhole
throat/bounce. Our strategy is straightforward:

(1) Take any $(3+1)$--dimensional hypervolume, and look for a
$3$--dimensional edgeless achronal spacelike hypersurface of strictly
minimal volume. Define such a surface, if it exists, to be the bounce
of a Tolman wormhole. This generalizes the Morris--Thorne flare out
condition for static traversable wormholes to arbitrary Tolman
wormholes.

(2) Use the Gauss--Codazzi and Gauss--Weingarten equations to
decompose the $(3+1)$--dimensional spacetime curvature tensor in terms
of the $3$--dimensional curvature tensor of the bounce and the
extrinsic curvature of the bounce as an embedded hypersurface in the
$(3+1)$--dimensional geometry.

(3) Reassemble the pieces: Write the spacetime curvature in terms of
the $3$--curvature of the bounce, and the extrinsic curvature of the
bounce in $(3+1)$ spacetime.

(4) Use the generalized flare-out condition to place constraints on
the stress-energy tensor at and near the throat.

A somewhat different but complementary strategy which dispenses
with the minimal volume condition in $(1)$ is then presented which
makes use instead of local properties of timelike geodesic congruences
near the candidate bounce. For this we replace $(1)$ by

$(1')$ The bounce of a Tolman wormhole is a 3-dimensional
spacelike hypersurface on which the expansion of a hypersurface
orthogonal timelike geodesic congruence vanishes identically and for
which the expansion is strictly positive to the immediate future of
the bounce and strictly negative to the immediate past.

This latter characterization in terms of geodesic expansion is useful
for when the volume of the hypersurface is ill-defined and is
equivalent to the latter definition when the volume integral
exists. {\em This version of the definition is also capable of dealing
with situations where only a part of the universe is ``bouncing''
while the rest continues its collapse, or is already in its expanding
phase. } One can deduce immediately the violation of the SEC in the
neighborhood of the bounce without having to follow steps $(2)-(4)$.
However, the analysis implied by these additional steps is crucial for
assessing the status of the other energy conditions (NEC, WEC, DEC) at
and near the bounce.

\section{Definition of a generic bounce}

We define a bounce, $\Sigma$, to be an edgeless achronal
$3$--dimensional spacelike hypersurface of {\em minimal}
volume. Compute the volume by taking
\begin{equation}\label{volume}
V(\Sigma) = \int \sqrt{{}^{(3)}g} \; d^{3} x.
\end{equation}
Now use Gaussian normal coordinates, $x^i=(\tau; \vec x^i)$, wherein the
hypersurface $\Sigma$ is taken to lie at $\tau=0$, so that
\begin{equation}
{}^{(3+1)}g_{\mu\nu} \; dx^\mu dx^\nu  = 
- d\tau^2 + {}^{(3)}g_{ij} \; dx^i dx^j.
\end{equation}
We do {\em not} demand that the manifold be globally of this form, but
will remain satisfied with the knowledge that such a coordinate system
exists and covers some open region surrounding the bounce.  The
variation in volume, obtained by pushing the hypersurface surface
$\tau=0$ out to $\tau = \delta \tau(x)$, is given by the standard
computation
\begin{equation}
\delta V(\Sigma) = 
\int {\partial\sqrt{{}^{(3)}g}\over \partial \tau} \; 
\delta \tau(x) \; d^{3} x.
\end{equation}
Which implies
\begin{equation}
\delta V(\Sigma) =
\int \sqrt{{}^{(3)}g} \;
{1\over2} \; g^{ij} \; {\partial g_{ij} \over \partial \tau} \; 
\delta \tau(x) \; d^{3} x.
\end{equation}
In Gaussian normal coordinates the extrinsic curvature is simply
defined by
\begin{equation}\label{extrinsic}
K_{ij} = - {1\over2} {\partial g_{ij} \over \partial \tau}.
\end{equation}
(See \cite[page~552]{MTW}. In this section we use MTW sign
conventions. The convention in \cite[page~156]{Visser} is opposite.)
Thus
\begin{equation}
\delta V(\Sigma) =  - \int \sqrt{{}^{(3)}g} \; \tr(K) \; 
\delta \tau(x) \; d^{3} x.
\end{equation}
[We use the notation $\tr(X)$ to denote $g^{ij} \; X_{ij}$.] Since
this is to vanish for arbitrary $\delta \tau(x)$, the condition that
the area be {\em extremal} is simply $\tr(K)=0$.  To force the volume
to be {\em minimal} requires (at the very least) the additional
constraint $\delta^2 V(\Sigma) \geq 0$. (We shall also consider
higher-order constraints below.)  But by explicit calculation
\begin{eqnarray}
\label{E:V2}
\delta^2 V(\Sigma) &=&  
- \int \sqrt{{}^{(3)}g} \;
\left( {\partial\tr(K)\over \partial \tau} - \tr(K)^2 \right) \; 
\nonumber\\
&&\qquad\qquad
\delta \tau(x) \; \delta \tau(x) \; d^{3} x.
\end{eqnarray}
Extremality [$\tr(K)=0$] reduces this minimality constraint to
\begin{equation}
\delta^2 V(\Sigma) =  
- \int \sqrt{{}^{(3)}g} \;
\left( {\partial\tr(K)\over \partial \tau} \right) \; 
\delta \tau(x) \; \delta \tau(x) \; d^{3} x \geq 0.
\end{equation}
Since this is to hold for arbitrary $\delta \tau(x)$ this implies that
at the bounce we certainly require
\begin{equation}
\label{E:simple}
{\partial\tr(K)\over \partial \tau} \leq 0.
\end{equation}
This is the simplest generalization of the ``flare-out'' condition for
FRW--based Tolman wormholes to arbitrary Tolman wormholes~\cite{Bounce}.
This simple bounce condition can be rephrased as follows: We have as
an identity that
\begin{equation}
\label{E:identity}
{\partial\tr(K)\over \partial \tau} = 
\tr\left({\partial K\over \partial \tau}\right) + 2 \tr(K^2).
\end{equation}
So minimality implies
\begin{equation}
\label{E:simple2}
\tr\left({\partial K\over \partial \tau}\right) + 2 \tr(K^2) \leq 0.
\end{equation}

We must now discuss some technical complications related to the fact
that we eventually prefer to have a strong inequality $(<)$ at or near
the bounce, than to have a weak inequality $(\leq)$ at the bounce
itself.  Similar technical complications arises when considering the
Morris--Thorne static spherically symmetric
wormhole~\cite{Morris-Thorne}, and the FRW-based Tolman wormholes
of~\cite{Bounce}.  These technical issues are also the main stumbling
block in setting up the analysis of generic traversable wormholes as
carried out in~\cite{HV:1,HV:2,HV:3,HV:4}. Unfortunately the details
are a little different for Tolman wormholes so we cannot simply copy
the previous arguments.

To set the notation, let us consider some one-parameter set of
deformations of the surface $\Sigma$ specified by
\begin{equation}
\delta \tau(x) = \epsilon f(x).
\end{equation}
This allows us to define a stratified collection of hypersurfaces
$\Sigma_\epsilon$ by taking
\begin{equation}
\Sigma_\epsilon = \left\{ \epsilon f(x), x^i \right\}.
\end{equation}
We now ask that, for all $f(x)$, the volume of these sets of
hypersurfaces $V[\Sigma_\epsilon]$ be a strict minimum at the
bounce. This is equivalent to asserting that for every ``direction''
$f(x)$ timelike deformations of the bounce lead to strict increases in
spatial volume. Now demanding that there is an open interval for which
$V[\Sigma_\epsilon] > V[\Sigma_0]$ leads, by the fundamental theorem
of calculus, to the existence of an open interval
\begin{equation}
\label{E:strict-volume}
\exists \; \tilde \epsilon>0: 
\forall \epsilon \in (-\tilde \epsilon,0)\cup(0,\tilde \epsilon) 
\qquad {d^2 V[\Sigma_\epsilon]\over d\epsilon^2} > 0.
\end{equation}
This then implies, via equation (\ref{E:V2})
\begin{eqnarray}
\label{E:strict-integrated}
&&
\exists \; \tilde \epsilon>0: 
\forall \epsilon \in (-\tilde \epsilon,0)\cup(0,\tilde \epsilon) 
\nonumber\\
&&
\qquad \int_{\Sigma_\epsilon} \sqrt{{}^{(3)}g} \; f^2(x)
\left( {\partial\tr(K)\over \partial \tau} - \tr(K)^2 \right) \; 
d^{3} x < 0.
\end{eqnarray}
Since this integral is negative for all $f(x)$ there will be some
$(3+1)$--dimensional open set ${\cal S}$ surrounding (but not
necessarily including) the bounce $\Sigma$ such that
\begin{eqnarray}
\label{E:strict-density}
&&
\left( {\partial\tr(K)\over \partial \tau} - \tr(K)^2 \right) \; < 0.
\end{eqnarray}
But we also know that $\tr(K)=0$ at the bounce itself. This allows us
to apply the fundamental theorem of calculus a second time to derive
the existence of a second open set $\tilde{\cal S}$ surrounding (but
not necessarily including) the bounce $\Sigma$ such that
\begin{eqnarray}
\label{E:strict}
&&
{\partial\tr(K)\over \partial \tau}  < 0.
\end{eqnarray}
To see this note that (\ref{E:strict-density}) can be written as
$dF(\tau)/d\tau - F(\tau)^2 <0$ on $\tau\in(0,\tau_*)$ with $F(0)=0$,
from which we see that $F(\tau)$ must initially go negative.  It is
this final version of the bounce condition that will lead to the most
general and powerful theorems.

These constraints on the extrinsic curvature lead to constraints on
the spacetime geometry, and consequently constraints on the
stress-energy tensor.

\section{Geometry at and near a generic bounce}

Using Gaussian normal coordinates in the region surrounding the bounce
the Gauss--Codazzi and Gauss--Weingarten equations give:
\begin{equation}
\label{E:Riemann-ijkl}
{}^{(3+1)}R_{ijkl} = {}^{(3)}R_{ijkl} + (K_{ik} K_{jl} - K_{il} K_{jk} ),
\end{equation}
\begin{equation}
\label{E:Riemann-tijk}
{}^{(3+1)}R_{\tau ijk} =  - (K_{ij|k}  - K_{ik|j}  ),
\end{equation}
\begin{equation}
\label{E:Riemann-titj}
{}^{(3+1)}R_{\tau i \tau j} =  {\partial K_{ij} \over \partial \tau}  + 
(K^2)_{ij}.
\end{equation}
See \cite[page~514~equations~(21.75)~and~(21.76)]{MTW} and
\cite[page~516~equation~(21.82)]{MTW}. Here the index $\tau$ refers to
the temporal direction normal to the three-dimensional bounce. As
usual, the vertical bar denotes a three-dimensional covariant
derivative built out of the three-dimensional spatial metric.

These results hold both on the throat and in the region
surrounding the throat: as long as the Gaussian
normal coordinate system does not break down. (Such breakdown being
driven by the fact that the normal geodesics typically intersect after
a certain distance.) 

Taking suitable contractions, and being careful {\em not} to use the
extremality condition $\tr(K)=0$, we find that {\em at and near} the
bounce:
\begin{eqnarray}
\label{E:Ricci-ij}
{}^{(3+1)}R_{ij} &=& 
{}^{(3)}R_{ij} 
- \left[ {\partial K_{ij} \over \partial \tau}  
+ 2 (K^2)_{ij} - \tr(K) \; K_{ij} \right],
\end{eqnarray}
\begin{eqnarray}
\label{E:Ricci-ti}
{}^{(3+1)}R_{\tau i} &=&   \tr(K)_{|i} -K_{ij}{}^{|j},
\end{eqnarray}
\begin{eqnarray}
\label{E:Ricci-tt}
{}^{(3+1)}R_{\tau \tau} &=&  
\tr\left({\partial K \over \partial \tau}\right)  + \tr(K^2)
\nonumber\\
&=& {\partial\tr(K)\over\partial \tau} - \tr(K^2).
\end{eqnarray}
So that the Ricci scalar is
\begin{eqnarray}
\label{E:Ricci-s}
{}^{(3+1)}R &=&  
{}^{(3)}R 
- \bigg[ 
2 \left({\partial \tr(K) \over \partial \tau} - \tr(K^2) \right)
\nonumber\\
&&  
\qquad
+ \tr(K^2) - \tr(K)^2 
\bigg].
\end{eqnarray}
To effect these contractions, we make use of the decomposition of
the spacetime metric in terms of the bounce  3-metric and the
set of three vectors ${e^{\mu}_i}$ tangent to the bounce and the
four-vector $n^\nu$ normal to the bounce:
\begin{equation}
{}^{(3+1)}g^{\mu\nu} = 
 - n^\mu \; n^\nu + e^{\mu}_i \; e^{\nu}_j \;\; {}^{(3)}g^{ij}.
\end{equation}
(Note the minus sign in front of the $n^\mu n^\nu$ term.)  For the
spacetime Einstein tensor ({\em cf.}
\cite[page~515~equations~(21.77)~and~(21.80)]{MTW} and
\cite[page~552~equations~(21.162a)--(21.162c)]{MTW}):
\begin{eqnarray}
\label{E:Einstein-ij}
{}^{(3+1)}G_{ij} &=&
{}^{(3)}G_{ij}  - \bigg[
{\partial K_{ij} \over \partial \tau}  
- g_{ij} {\partial \tr(K) \over \partial \tau}  
- \tr(K) K_{ij} 
\nonumber\\
&&
+ 2 (K^2)_{ij}
+ {1\over2} g_{ij} \; \left[\tr(K^2)+\tr(K)^2\right] \bigg].
\end{eqnarray}
\begin{eqnarray}
\label{E:Einstein-ti}
{}^{(3+1)}G_{\tau i} &=&   
\tr(K)_{|i} -K_{ij}{}^{|j}.
\end{eqnarray}
\begin{eqnarray}
\label{E:Einstein-tt}
{}^{(3+1)} G_{\tau \tau} &=& 
+ {1\over2} {}^{(3)} R  - {1\over2} \left[ \tr(K^2) - \tr(K)^2 \right].
\end{eqnarray}
The calculations presented above are simply a matter of brute force
index gymnastics---but we feel that there are times when explicit
expressions of this type are useful.

\section{Constraints on the stress-energy tensor}

\subsection{First constraint: SEC violation}

By using the Einstein equations, $G_{\mu\nu} = 8\pi G\; T_{\mu\nu}$,
the SEC applied to the stress-energy tensor is equivalent to the Ricci
convergence condition~\cite{Visser}:
\begin{equation}
\forall \hbox{ timelike } V^\mu :  
\qquad R_{\mu\nu} \; V^\mu V^\nu > 0.
\end{equation}
But by the simple flare-out condition (\ref{E:simple}), and equation
(\ref{E:Ricci-tt}), we see ${}^{(3+1)}R_{\tau \tau} \leq 0$. This implies
that the SEC is either violated or on the verge of being violated at
the throat. To really pin down SEC violation we must invoke the
stricter inequality (\ref{E:strict}) to see that the SEC is definitely
violated in some open region surrounding the bounce.

Equivalently, the spacetime Ricci tensor ${}^{(3+1)}R_{\mu\nu}$ has at
least one negative definite eigenvalue (corresponding to a timelike
eigenvector) everywhere in some open region surrounding the bounce.  A
similar result for Euclidean wormholes is quoted
in~\cite{Giddings-Strominger} and the present analysis can of course
be carried over to Euclidean signature with appropriate definitional
changes.

\subsection{Second constraint: density}

The energy density in the vicinity of the bounce is
\begin{equation}
\label{E:density}
\rho \equiv T_{\tau\tau} = {1\over8\pi G} G_{\tau\tau} 
= {1\over16\pi G}
\left[
{}^{(3)} R  -  \tr(K^2)  + \tr(K)^2
\right].
\end{equation}
The above is the generalization of the result that for a FRW--based
Tolman wormhole~\cite{Bounce}
\begin{equation}
\rho= {3 \over8\pi G}\left[{k\over a^2} + {\dot a^2\over a^2} \right].
\end{equation}
(With MTW conventions ${}^{(3)} R = 6/a^2$ for a three-sphere.) Since
$\tr(K)=0$ at the bounce, we see that at the bounce itself
\begin{equation}
\rho_{\mathrm bounce} 
\leq {1\over16\pi G}
\left[
{}^{(3)} R
\right].
\end{equation}
Thus a {\em necessary} condition for the energy density to be positive
at the bounce is that the bounce be a three-manifold of everywhere
positive Ricci scalar.

\subsection{Third constraint: average pressure}

Define an average pressure by
\begin{eqnarray}
p &\equiv& {1\over3} \;  g^{ij} \;\; {}^{(3+1)}T_{ij} =
{1\over24\pi G} \; \; g^{ij} \;\; {}^{(3+1)}G_{ij}.
\end{eqnarray}
Then
\begin{eqnarray}
\label{E:pressure}
p
&=&
{1\over16\pi G}  
\Bigg[ 
-{1\over3} \; {}^{(3)}R + {1\over3} \left[ \tr(K^2) - \tr(K)^2 \right]
\nonumber\\
&&
\qquad
+ {4\over3} \left[ {\partial \tr(K) \over \partial \tau} - \tr(K^2) \right]
\Bigg].
\end{eqnarray}
The above is the generalization of the result that for a FRW--based
Tolman wormhole~\cite{Bounce}
\begin{equation}
p= -{1 \over8\pi G}\left[{k\over a^2} 
+ {\dot a^2\over a^2} + 2{\ddot a\over a}\right].
\end{equation}
Now at and near the bounce we can write the average pressure as
\begin{eqnarray}
p
&=&
-{1\over3} \; \rho +
{1\over12\pi G}  
\left[ {\partial \tr(K) \over \partial \tau} - \tr(K^2) \right].
\end{eqnarray}
The term in square brackets is negative definite by (\ref{E:strict}),
so there is an open region surrounding the bounce for which
\begin{eqnarray}
p &<& -{1\over3} \; \rho.
\end{eqnarray}
This is just the previously discussed SEC violation in another
disguise, though it has the advantage of emphasizing the fact that
positive densities near the bounce imply negative pressures near the
bounce.

\subsection{Fourth constraint: energy conditions}

Using the average pressure defined above, it is easy to prove that,
even in the absence of any symmetries
\begin{equation}
\hbox{NEC} \implies \quad 
(\rho + p \geq 0 ).
\end{equation}
\begin{equation}
\hbox{WEC} \implies \quad 
(\rho \geq 0 ) \hbox{ and } (\rho + p \geq 0).
\end{equation}
\begin{equation}
\hbox{SEC} \implies \quad 
(\rho + 3 p \geq 0 ) \hbox{ and } (\rho + p \geq 0).
\end{equation}
\begin{equation}
\hbox{DEC} \implies \quad 
(\rho \geq 0 ) \hbox{ and } (\rho \pm p \geq 0).
\end{equation}
Basic definitions of the energy conditions are given
in~\cite{Visser,Hawking-Ellis}. It is important to note that in the
case of a FRW universe these implications ($\implies$) are
strengthened to equivalences ($\iff$) as discussed
in~\cite{Galaxy,Galaxy-2,Galaxy-3}.

To see how these relations are proved, focus as an example on the NEC,
which states that for all null vectors $T_{\mu\nu} \; V^\mu V^\nu \geq
0$. Note that (up to arbitrary normalization) all null vectors can be
written $V^\mu = (1; \beta^i)$ with $g_{ij} \; \beta^i \beta^j =
1$. Therefore, for all $\beta^i$ we have
\begin{equation}
\rho + 2 \; f_i \; \beta^i + T_{ij} \; \beta^i \beta^j \geq 0,
\end{equation}
where the momentum flux is defined by $f_i = T_{\tau i}$.  By
averaging over the two null vectors $(1; \beta^i)$ and $(1; -\beta^i)$
this implies that for all $\beta^i$
\begin{equation}
\rho + T_{ij} \; \beta^i \beta^j \geq 0.
\end{equation}
Finally average over three mutually perpendicular unit vectors
$\beta^i$
\begin{equation}
\rho + {1\over3} T_{ij} \; g^{ij} \geq 0.
\end{equation}
Equivalently
\begin{equation}
\rho + p \geq 0.
\end{equation}
The same logic can now be followed for the other pointwise energy
conditions.

It therefore  becomes interesting to use the Einstein equations to
calculate $\rho\pm p$. We find
\begin{eqnarray}
\label{E:rho+p}
\rho+p &=& {1\over16\pi G}  
\Bigg[ 
{2\over3} \; {}^{(3)}R - {2\over3} \left[ \tr(K^2) - \tr(K)^2 \right]
\nonumber\\
&&
\qquad
+ {4\over3} \left[ {\partial \tr(K) \over \partial \tau} - \tr(K^2) \right]
\Bigg],
\nonumber\\
&&
\end{eqnarray}
and
\begin{eqnarray}
\label{E:rho-p}
\rho-p &=& {1\over16\pi G}  
\Bigg[ 
{4\over3} \; {}^{(3)}R - {4\over3} \left[ \tr(K^2) - \tr(K)^2 \right]
\nonumber\\
&&
\qquad
- {4\over3} \left[ {\partial \tr(K) \over \partial \tau} - \tr(K^2) \right]
\Bigg].
\nonumber\\
&&
\end{eqnarray}
We shall now show that there is an enormous class of spacetime
geometries for which these two quantities are positive at and near the
bounce. To see this, consider the following scaling argument: suppose
we have some spacetime geometry which has a bounce, and for which the
bounce is a manifold of positive Ricci scalar. Now consider the class
of geometries
\[
g\to g_\epsilon: ds^2 = -dt^2 + \epsilon^2 g_{ij} \; dx^i dx^j.
\]
For this class of geometries
\[
{}^{(3)}R\to {}^{(3)}R_\epsilon = {{}^{(3)}R\over\epsilon^2},
\]
while on the other hand $\tr(K)$ and $\tr(K^2)$ are independent of
$\epsilon$. [$K_{ij}\to \epsilon^2 K_{ij}$ 
but $g^{-1} \to \epsilon^{-2} g^{-1}$,
so $ \tr(K)\to \tr(K)$. ] Thus for $\epsilon$ sufficiently small the
intrinsic curvature terms will always dominate over the extrinsic
curvature terms and we can guarantee that the density [equation
(\ref{E:density})] and equations (\ref{E:rho+p})--(\ref{E:rho-p}) are
all positive.  Thus there is a large class of bounce geometries that
are compatible with the NEC, WEC, and DEC. However bounce geometries
must always violate SEC. This generalizes the result for FRW--based
Tolman wormholes presented in~\cite{Bounce}.  Somewhat stronger
statements can be made by looking at the explicit formulae for the
components of the Einstein tensor
\begin{eqnarray}
\label{E:Einstein-ij+}
{}^{(3+1)}G_{ij}(\epsilon) &=&
{}^{(3)}G_{ij} \; \epsilon^{-2} + O(\epsilon^2).
\end{eqnarray}
\begin{eqnarray}
\label{E:Einstein-ti+}
{}^{(3+1)}G_{\tau i} (\epsilon)&=&  O(1).
\end{eqnarray}
\begin{eqnarray}
\label{E:Einstein-tt+}
{}^{(3+1)} G_{\tau \tau}(\epsilon) &=& 
+ {1\over2} \; {}^{(3)} R \; \epsilon^{-2} + O(1).
\end{eqnarray}
By choosing $\epsilon$ small enough we can guarantee that NEC, WEC,
and DEC are satisfied, though SEC must always be violated.

\section{Generic Bounces Defined using Timelike Geodesics}

The definition of a generic bounce starting from the volume integral
in (\ref{volume}) is similar in spirit to and motivated by the
definition of a generic wormhole throat developed in \cite{HV:1,HV:2},
but there are important differences we would like to underscore. First
of course, is the fact that a bounce is by definition an intrinsically
time-dependent phenomena, whereas wormholes may be either static or
time-dependent. Second, whereas wormhole throats in spacetime are
defined via two-dimensional spacelike hypersurfaces, the bounce is a
three-dimensional spacelike hypersurface. Third, and perhaps the most
important difference stems from the fact that whereas wormhole throats
are always closed (and thus have finite area) spatial hypersurfaces
satisfying certain extremality and minimality properties, bounces may
be spatially open (e.g., as in a FRW cosmology with flat or hyperbolic
spatial sections) or closed (e.g., as in a FRW cosmology with closed
spatial sections), depending on the type of cosmology being
considered.  In the latter case, the spatial volume integral is of
course finite and well defined, but in the former case, it is not
finite, and a definition of a generic bounce is called for which is
not bound up with potentially infinite integrals, but which is
nevertheless fully equivalent to the definition given earlier in this
paper. That such a {\it local} pointwise definition of a generic
bounce is possible is strongly suggested by the work in \cite{HV:3}
and \cite{HV:4}, which treated general dynamic wormholes on the basis
of (null) geodesic congruences.  The idea is simply to define what we
mean by a bounce in terms of the local properties of timelike geodesic
congruences in the neighborhood of the putative bounce. This is
motivated by the very physical question which asks how is the motion
of test particles in the vicinity of a bounce affected by that bounce?
The alternative definition is as follows: a bounce is a 3-dimensional
spatial hypersurface such that the timelike geodesic congruence
orthogonal to it vanishes on the hypersurface, is strictly expanding
to the immediate future of the hypersurface, and is strictly
contracting to the immediate past. This definition is capable of
dealing with situations where only {\em part} of the universe is
``bouncing'', while the rest either continues its collapse or is
already in an expanding phase. The vanishing condition is equivalent
to the minimality condition obtained in Section II and the
contraction-expansion condition is none other than the Morris-Thorne
``flare-out'' condition generalized to bounces.  Indeed, the mutual
spreading out of a ``swarm'' of future-directed test particles in the
immediate future of the bounce is what we mean by ``flare-out''.  As
we will see, all these notions are pointwise.  Our next task is to
make them precise. In this section, we follow the same sign
conventions and notation as used in \cite{HV:3,HV:4} which are taken
from Wald \cite{Wald}.

So, consider a timelike geodesic congruence orthogonal to the spatial
hypersurface $\Sigma$, to be conveniently located without loss of
generality at $\tau = 0$, and let $\xi^a$ denote a tangent vector to a
geodesic in this congruence; we can always arrange for all these
tangents, parameterized by proper time $\tau$, to have identical
normalization:
\begin{equation}
\xi^a \xi_ a = g_{ab} \; \xi^a \xi^b = -1,
\end{equation}
where the spatial and spacetime metrics are related by
\begin{equation}
^{(3)}g_{ab} = {^{(3+1)}}g_{ab} + \xi^a \xi^b.
\end{equation}
Now define the tensor field
\begin{equation}\label{K}
K_{ab} \equiv \nabla_b \xi_a,
\end{equation}
by using the normalization condition and the fact that tangent vectors
are parallel transported $(\xi^a \nabla_a \xi^b = 0)$, one can easily
show that this tensor is purely spatial, i.e., $\xi^a K_{ab} = \xi^b
K_{ab} = 0$, and moreover is symmetric, $K_{ab} = K_{ba}$, because the
congruence is hypersurface orthogonal. This tensor is in fact the
extrinsic curvature of the hypersurface $\Sigma$ and measures the
``degree of bending'' with respect to the embedding spacetime, as is
well known. But it also contains useful information regarding the
expansion $\theta$ and the traceless shear $\sigma_{ab}$
\begin{eqnarray}
\theta &=& ^{(3)}g_{ab} K^{ab} = {\rm tr}(K) ,\\
\sigma_{ab} &=& K_{(ab)} - \frac{1}{3} \; {^{(3)}g_{ab}} \, \theta,
\end{eqnarray}
of the timelike geodesic congruence normal to the hypersurface. The
expansion $\theta$ measures the instantaneous ``spreading'' or
divergence of nearby timelike geodesics while the symmetric shear
tensor measures the ``slippage'' of nearby geodesics. The shear is a
purely spatial tensor, which immediately implies that $\sigma^{ab}
\sigma_{ab} \geq 0$, is always a positive semi-definite quantity.

The rates of change of the expansion and shear with respect to proper
time ($\tau$ of the test particles) can be calculated, and in the case
of the expansion, one obtains a simplified version of the celebrated
Raychaudhuri equation \cite{Wald}
\begin{equation}
\label{Ray}
\frac{d \theta}{d \tau}= -\frac{1}{3} \theta^2 -\sigma_{ab}\sigma^{ab}
-R_{ab} \; \xi^a \, \xi^b,
\end{equation}
where $R_{ab}$ is the Ricci tensor of the full spacetime.  This is
independent of coordinate system.  This is a simplified version
because the additional contribution from the twist, or anti-symmetric
part of $K_{ab}$ is absent here, since we are dealing with a
hypersurface orthogonal congruence.  Note that equation
(\ref{E:Ricci-tt}) is actually this special case Raychaudhuri equation
(\ref{Ray}) in disguise [once we express equation (\ref{Ray}) in terms
of Gaussian coordinates and take into account the relative sign in the
definitions for the extrinsic curvature used in this Section, equation
(\ref{K}) and in Section II, equation (\ref{extrinsic})].

With these simple preliminaries out of the way, we can now give the
(local) definition of what it means to be a generic bounce. A bounce
is any three-dimensional spatial hypersurface on which the expansion
of a hypersurface orthogonal timelike geodesic congruence vanishes
identically:
\begin{equation}
(i) \qquad 
\theta(0) = 0,
\end{equation}
and for which the expansion is positive to the immediate future
and negative to the immediate past:
\begin{equation}
(ii) \qquad 
\exists \; \tilde \tau_+ >0: 
\forall \tau \in (0,\tilde \tau_+), 
\qquad
\theta(\tau) > 0;
\end{equation}
\begin{equation}
(iii) \qquad
\exists \; \tilde \tau_- >0: 
\forall \tau \in (-\tilde \tau_-,0), 
\qquad
\theta(\tau)< 0.
\end{equation}
These three properties of the timelike geodesics capture the
minimality and flare-out conditions of a bounce directly without
needing to refer to the volume of the bounce.  Indeed, a bunch of test
particles traversing the bounce will initially have a cross section
that first decreases in time, reaching a minimum at the throat,
followed by a subsequent increase.  A similar characterization was
successfully employed recently in defining the general time-dependent
wormhole throat \cite{HV:3,HV:4}, by means of null geodesics. By the
fundamental theorem of calculus conditions (i), (ii), and (iii) can be
combined to imply
\begin{equation}
\label{E:combined}
\exists \; \tilde \tau_0 >0: 
\forall \tau \in (-\tilde \tau_0,0)\cup(0,\tilde \tau_0), 
\qquad
{d\theta\over d\tau} > 0.
\end{equation}

It will be noted that the Raychaudhuri equation (\ref{Ray}) is
independent of the underlying dynamics of the geometry: it is a
statement only about (timelike) geodesics (test particles) in a
particular geometry.  If we {\it now} impose Einstein's equation (the
geometrodynamics)
\begin{equation}
R_{ab} = 8\pi G\; \Big(T_{ab} - \frac{1}{2}g_{ab} T \Big),
\end{equation}
and make use of the three conditions (i), (ii), and (iii) [in the
form of equation (\ref{E:combined})] then by the Raychaudhuri equation
we must conclude that
\begin{eqnarray}
&&\exists \; \tilde \tau_0 >0: 
\forall \tau \in (-\tilde \tau_0,0)\cup(0,\tilde \tau_0), 
\nonumber\\
&& \qquad
\xi^a \xi^b \Big( T_{ab} - \frac{1}{2}g_{ab} T \Big)< 0.
\end{eqnarray} 
That is, the SEC is strictly violated in an open region
surrounding the bounce.

\section{Discussion}

One of the key results of traversable wormhole physics, perhaps {\em
the} key result, is the unavoidable violations of the null energy
condition (NEC) at or near the throat
\cite{Morris-Thorne,MTY,Visser,HV:1,HV:2,HV:3,HV:4}.  In the case of a
Tolman wormhole it is instead the strong energy condition (SEC) that
is violated at or near the bounce~\cite{Bounce}.  We have developed a
number of general theorems that characterize the extent and generality
of these SEC violations. An important point is that it is relatively
easy to obtain SEC violations, they can be found already at the
classical level and do not even require the standard appeal to quantum
effects that is common in seeking to justify NEC
violations~\cite{Visser:ANEC}.

There are a number of powerful constraints that can be placed on the
stress-energy tensor at and near the bounce of a Tolman wormhole
simply by invoking the minimality properties of the bounce. Depending
on the precise form of the assumed flare-out condition, these
constraints give the various energy condition violation theorems we
are seeking. Even under the weakest assumptions they constrain the
stress-energy tensor to at best be on the verge of violating the SEC.

In this article we have sought to give an overview of the energy
condition violations that occur in generic Tolman wormholes. We point
out that these violations of the energy conditions follow unavoidably
from the definition of a Tolman wormhole (bounce) and the definition
of the total stress-energy tensor via the Einstein equations.
To show the generality of the energy condition violations, we have
developed an analysis that is capable of dealing with Tolman wormholes
of arbitrary symmetry.  We have presented two complementary
definitions of a bounce that agree where they overlap but are much
more general than the FRW--based bounces considered
in~\cite{Bounce}.  The present definitions work well in any spacetime
and nicely capture the essence of the idea of what we would want to
call a Tolman wormhole.  We do not need to make any assumptions about
the existence of any asymptotic regions, nor do we need to assume that
the manifold is topologically non-trivial. It is important to realize
that the essence of the definitions lie in the local geometrical
structure of the bounce.

In the broader scheme of things, this article should be viewed as a
contribution to the continuing debate as to whether the universe
emerged form a mathematical singularity in the big bang, or if
something more subtle is going on. While there can be little doubt
that the universe emerged from a hot dense fireball colloquially
called the big bang, it is a big step from a hot dense fireball to a
mathematical singularity. For many years it was believed that the
Penrose~\cite{Hawking-Ellis} and Geroch~\cite{Wald} cosmological
singularity theorems definitively proved the existence of a
mathematical singularity, but these theorems are based on assuming the
SEC. This article demonstrates that these theorems {\em cannot} be
improved in the sense that we have exhibited a large class of Tolman
wormholes that satisfy all energy conditions {\em except} the
SEC. Furthermore, there is now a large body of evidence pointing to
the fact that the SEC may not be the fundamental physical restriction
it was once thought to be: there are many quite reasonable physical
systems, even classical systems, that violate the SEC
\cite{Bounce,Bekenstein,Parker-Fulling,Parker-Wang,Rose:86,Rose:87,Kandrup}.
Likewise, gravitational vacuum polarization, although it is a small
quantum effect, often violates the SEC (and other energy
conditions)~\cite{Visser:ANEC,gvp1,gvp2,gvp3,gvp4}.

As discussed in~\cite{Bounce} there are a number of singularity
theorems provable within the ``eternal inflation''
paradigm~\cite{Borde94a,Borde94b,Borde94c,Borde96,Borde97}, but these
theorems obtain their results at the cost of making rather specific
additional hypotheses and they are not in conflict with the results of
the present paper.

Finally we should mention that a particularly large class of quite
reasonably behaved Tolman wormholes is provided by the analytic
continuation of Euclidean wormholes back to Lorentzian
signature~\cite{Coule}.

\section*{Acknowledgments}

This work was supported in part by the US Department of Energy
(MV and CMP) and by the Spanish Ministry of Culture and Education
(DH).  Matt Visser also wishes to thank LAEFF (Laboratorio de
Astrof\'\i{}sica Espacial y F\'\i{}sica Fundamental; Madrid, Spain),
and Victoria University (Te Whare Wananga o te Upoko o te Ika a Maui;
Wellington, New Zealand) for hospitality during final stages of this
work.



\end{document}